\documentclass[sigconf]{acmart}
\usepackage{booktabs} 
\usepackage{soul}
\usepackage{bm}
\usepackage{algorithm}
\usepackage{algpseudocode}
\usepackage{comment}
\usepackage{verbatim}
\usepackage{hyperref}
\usepackage{fancybox}
\usepackage{balance}
\usepackage{graphicx}
\usepackage{epstopdf}
\newcommand{\hypobox}[1]{

        \begin{center}\noindent\thicklines\setlength{\fboxsep}{8pt}\cornersize{0.2}\ovalbox{

                \begin{minipage}{3.0in}

                        \textit{#1}

                \end{minipage}} 

        \end{center}}

\usepackage{color, colortbl}
\definecolor{Gray}{gray}{0.9}

\usepackage{ifthen}
\usepackage{amssymb}
\newboolean{showcomments}
\setboolean{showcomments}{true} 
\ifthenelse{\boolean{showcomments}}
  {\newcommand{\nb}[2]{
    \fcolorbox{gray}{yellow}{\bfseries\sffamily\scriptsize#1}
    {\sf\small$\blacktriangleright$\textit{#2}$\blacktriangleleft$}
   }
   
  }
  {\newcommand{\nb}[2]{}
   
  }

\newcommand{\todocite}[1]{\smash{\fcolorbox{gray}{red}{[?]}}}

\copyrightyear{2020}
\acmYear{2020}
\setcopyright{acmlicensed}
\acmConference[MSR '20]{17th International Conference on Mining Software Repositories}{October 5--6, 2020}{Seoul, Republic of Korea}
\acmBooktitle{17th International Conference on Mining Software Repositories (MSR '20), October 5--6, 2020, Seoul, Republic of Korea}
\acmPrice{15.00}
\acmDOI{10.1145/3379597.3387478}
\acmISBN{978-1-4503-7517-7/20/05}

\begin{document}
\title{Detecting and Characterizing Bots that Commit Code}

\author{Tapajit Dey}
\orcid{0000-0002-1379-8539}

\author{Sara Mousavi}
\affiliation{%
  \institution{The University of Tennessee}
  \streetaddress{1520 Middle Dr.}
  \city{Knoxville}
  \state{TN}
  \country{USA}
  \postcode{37996}
}
\email{tdey2@vols.utk.edu}

\email{mousavi@vols.utk.edu}

\author{Eduardo Ponce}
\orcid{0000-0002-8854-9043}

\author{Tanner Fry}
\affiliation{%
  \institution{The University of Tennessee}
  \streetaddress{1520 Middle Dr.}
  \city{Knoxville}
  \state{TN}
  \country{USA}
  \postcode{37996}
}
\email{eponcemo@utk.edu}
\email{tfry2@vols.utk.edu}

\author{Bogdan Vasilescu}
\affiliation{%
  \institution{Carnegie Mellon University}
  \streetaddress{Forbes Avenue}
  \city{Pittsburgh}
  \state{PA}
  \country{USA}
  \postcode{15213}
}
\email{vasilescu@cmu.edu}

\author{Anna Filippova}
\affiliation{%
  \institution{Github}
  \streetaddress{88 Colin P. Kelly Jr. St.}
  \city{San Francisco}
  \state{CA}
  \country{USA}
  \postcode{94107}
}
\email{annafil@github.com}

\author{Audris Mockus}
\affiliation{%
  \institution{The University of Tennessee}
  \streetaddress{1520 Middle Dr.}
  \city{Knoxville}
  \state{TN}
  \country{USA}
  \postcode{37996}
}
\email{audris@utk.edu}

\renewcommand{\shortauthors}{T. Dey et al.}

\begin{abstract}
  Background: Some developer activity traditionally performed
  manually, such as making code commits, opening, managing, or
  closing issues is increasingly subject to automation in many OSS
  projects. Specifically, such activity is often performed by tools that
  react to events or run at specific times. We refer to such
  automation tools as bots and, in many software mining scenarios
  related to developer productivity or code quality, it is desirable
  to identify bots in order to separate their actions from actions
  of individuals. 
  Aim: Find an automated way of identifying bots and
  code committed by these bots, and to characterize the types of
  bots based on their activity patterns.  
  Method and Result: We propose \textbf{BIMAN}, a systematic approach to detect bots using author names, commit messages, files modified by the commit, and projects associated with the commits. For our test data, the value for AUC-ROC was 0.9. We also characterized these bots based on the time
  patterns of their code commits and the types of files
  modified, and found that they primarily work with documentation files and web pages, and these files are most prevalent in HTML and JavaScript ecosystems. We have compiled a shareable dataset containing detailed
  information about 461 bots we found (all of which have more than 1000 commits) and 13,762,430 commits they created. 
\end{abstract}

%
%

\keywords{bots, automated commits, random forest, ensemble model, social coding platforms, software engineering}

\maketitle

\section{Introduction}\label{s:intro}
Bot is a classification assigned to a software application that performs automated tasks
based on a predefined set of instructions, and it either runs continuously or is triggered by events associated with events, time conditions, or manual execution.
Examples of applications that can function as bots are automated scripts,
activity loggers~\cite{frenot2012logos}, web crawlers~\cite{boldi2018bubing}, and chat bots~\cite{bradley2018context,gianvecchio2008measurement}.
A large number of software developers, teams, and companies use bots to do various, often repetitive, tasks, because bots can perform those tasks more efficiently than human users~\cite{Farooqintegration,lebeuf2017software,monperrus2019explainable}.

In social coding platforms~\cite{dabbish2012social,cosentino2017systematic} such as GitHub and BitBucket a number of bots regularly create code commits, issues, and pull requests. However, detecting a bot is a challenging task because on the surface there is no apparent difference between the activity of a bot and that of a human.
Moreover, the message structure, message content, and linguistic style of a code commit created by a bot can look very similar to a commit created by a human author.
While there are a number of well-known and active bots, such as 
Dependabot\footnote{\url{https://dependabot.com}} and
Greenkeeper,\footnote{\url{https://greenkeeper.io}} not all bots are as popular
and easily recognizable, as we disclose in this work.

Our review of the existing literature did not reveal any systematic 
approach for determining whether a given author in a social coding platform is 
a bot.
Therefore, in this work, we propose \textbf{BIMAN} --- \textit{Bot Identification by commit Message, commit Association, and author Name} --- a novel technique to detect bots that commit code. \textbf{BIMAN} is comprised of three methods that consider independent aspects of the commits made by a particular author: 1) \textit{Commit Message:} Identify if commit messages are being generated from templates; 2) \textit{Commit Association:} Predict if an author is a bot using a random forest model, with features related to files and projects associated with the commits as predictors; and
3) \textit{Author Name:} Match author's name and email to common bot patterns.
The code for \textbf{BIMAN} is available at our GitHub repository.~\footnote{\url{https://github.com/ssc-oscar/BIMAN_bot_detection}~\cite{tapajit_dey_2020_3711620}}
    
We applied \textbf{BIMAN} to the \textit{World of Code} dataset~\cite{woc19}, which has a collection of more than 34 million authors who have committed code to a GitHub repository, along with detailed information for approximately 1.6 billion commits made by these authors.
A \textbf{dataset} was compiled with information about 461 bots, detected by \textbf{BIMAN} and manually verified as bots, each with more than 1,000 commits, along with detailed information about 13,762,430 commits made by these bots. This dataset is available at \textcolor{red}{\url{https://zenodo.org/record/3694401}}~\cite{tapajit_dey_2020_3694401}.

We also aim to characterize the bots found using \textbf{BIMAN} based on their activity, that is, the type of files modified and the commit timestamp, which can provide insights into the type of work they perform and the programming languages they work with. 
We discovered four different classes of bots based on the pattern of their activity over the 24 hours of a day and identified the type of files commonly edited by bots.

In summary, we make the following contributions in this paper:
\textbf{1)} \textbf{BIMAN}, a generalizable technique combining three different methods for identifying if a given author is a bot;
\textbf{2)} Characterization of bots based on their activity patterns; and
\textbf{3)} A labeled dataset comprised of 461 bots with 13,762,430 commits.
We expect our efforts will be useful in enabling further research in software development requiring either the inclusion or exclusion of bots. 

The rest of the paper is organized as follows: We discuss the motivation for our work and the specific research questions addressed in this paper in Section~\ref{s:motiv}.
In Section~\ref{s:relwork}, we discuss related works in the topic. Section~\ref{s:method} focuses on the proposed methods for detecting and characterizing bots that commit code. In Section~\ref{s:result}, we describe the results we found pertaining to our research questions. Finally, we discuss the limitations of the current version of our work and the possible future works in Section~\ref{s:limit} and conclude the paper in Section~\ref{s:conclusion}.

\section{Motivation and Research Questions}\label{s:motiv}

The main motivation behind our bot detection effort is twofold: 1) Data cleaning: the automated nature of bots can significantly affect the estimates of team size, the amount of activity, and developer productivity, which can threaten the validity of such measures and any decision based on such measures; and 2) Research: enabling further research into bots.

Many software researchers look at the activity of software developers for understanding their cultural behavior~\cite{dey2019patterns,amreen2019methodology, dey2020pull,dey2018software,dey2018modeling,dey2020deriving}, estimating team size~\cite{bhowmik2015optimal}, measuring productivity~\cite{zhou2010developer}, and studying developer interaction such as knowledge flow within a project~\cite{mockus2009succession,kerzazi2016knowledge} and prediction of build failures~\cite{wolf2009predicting}. While conducting such studies, it is important to account for developers that are bots because bots typically have different activity patterns than humans. For example, bots may generate physically impossible metrics of activity and productivity, or could at least significantly bias these estimates.
Furthermore, the desire to stand out can lead to creation of extreme numbers of files or commits via automation (e.g., GitHub author \textit{one-million-repo}\footnote{\url{https://github.com/one-million-repo}} has 1,102,551 commits and the repository \textit{biggest-repo-ever}~\footnote{\url{https://github.com/one-million-repo/biggest-repo-ever}} has 9,960,000 commits).~\footnote{\url{https://bitbucket.org/swsc/overview/src/master/fun}}

However, the first step in adopting a data cleaning scheme to mitigate the effects of bots in software engineering research is to find the bots and, as mentioned earlier, we found no systematic approach for that. Therefore, the first research question we address in this paper is:

\textbf{RQ1: How can we determine if a particular author is a bot?}

A logical follow-up to this research question is characterizing the bots found. Previous methodologies proposed for bot characterization are to examine the design and construction of bots~\cite{erlenhov2019current}
and their intrinsic properties and interaction patterns~\cite{lebeuf2018taxonomy}, while we strive to characterize bots based on activity patterns, for example, type of files modified and commit timestamp.

In contrast to the existing taxonomy for bots, which is generated using a theoretical setting, (e.g., Erlenhov et. al. used faceted analysis~\cite{erlenhov2019current} and Lebeuf~\cite{lebeuf2018taxonomy} used the taxonomy generation method proposed by~\cite{usman2017taxonomies}), we select a data intensive technique to characterize bots using \textbf{BIMAN} because of the limited information available other than what can be obtained from their commit activity.

The general assumption about bots is that they primarily perform tedious tasks, and we want to investigate the veracity of this conjecture.
In addition, we want to estimate the prevalence of bots in relation to programming languages, which can highlight the type of work they perform and the areas with scope for bot adoption. Therefore, we pose our second research question as:

\textbf{RQ2: What type of work do bots perform and which programming languages do they work with?}

\section{Related Works}\label{s:relwork}

The idea of ``bots'', or software applications that can imitate human activity, dates back to 1950 with Alan Turing asking the question ``Can machines think?''~\cite{alan1950turing}. Recent advancements in artificial intelligence, especially natural language processing and machine learning have led to a proliferation of bots across domains, such as in virtual assistants 
(Apple's Siri~\cite{winarsky2012development} and Google's Assistant~\cite{statt2016google}, and Amazon's Alexa~\footnote{https://developer.amazon.com/en-US/alexa}),
education~\cite{kerry2008conversational,benotti2014engaging}, e-commerce~\cite{thomas2016business}, customer service~\cite{jain2018convey}, and social media platforms~\cite{abokhodair2015dissecting}. 

Open-source software (OSS) communities, and software engineering in general, primarily use bots to reduce the workload of repetitive tasks. Wessel et. al.~\cite{wessel2018power} studied 351 GitHub projects with more than 2,500 stars and found that 26\% of them use bots, with bot usage rising since 2013. Bots also support communication and decision making~\cite{storey2016disrupting,perez2017rise}, automate deployment and evaluation of software~\cite{beschastnikh2017accelerating}, and automate tasks that would require human interaction in collaborative software development platforms~\cite{lebeuf2017software}.

However, while these studies highlight how bots are used and how prevalent bot adoption is in popular OSS projects, they do not present any generalizable method to detect the bots that are already present. Wessel et. al.~\cite{wessel2018power} inspected the GitHub account of a suspected bot and checked if it is tagged as a bot. They also examined pull request messages, in search of obvious messages, for instance: ``This is an automated pull request to...''. Erlenhov et. al.~\cite{erlenhov2019current} and Lebeuf~\cite{lebeuf2018taxonomy} analyzed 11 and 3 well-known bots, respectively, and neither suggested a formal method of detecting if a given author is a bot.

In terms of bot characterization, Lebeuf~\cite{lebeuf2018taxonomy} proposed characterizing bots by analyzing 22 facets organized into 3 dimensions: Environmental (where the bot operates), Intrinsic (internal bot properties), and Interaction (how the bot interacts with its environment). Erlenhov et. al.~\cite{erlenhov2019current} focused on the ``DevBots'', i.e., bots that support software development, and proposed a taxonomy comprising 4 facets: Purpose (general or specialized), Initiation (triggered by users and/or system), Communication (how the bot communicates with other users), and Intelligence (adaptive or static). Dey et. al.~\cite{dey2020exploratory} used the dataset shared with this work to categorize bot commits by the type of file operations (add, delete, or modify), find the distribution of file types changed, and identify the file types that tend to get updated together.

\section{Methodology}\label{s:method}

\begin{figure*}[t]
\centering
\includegraphics[width=0.7\linewidth]{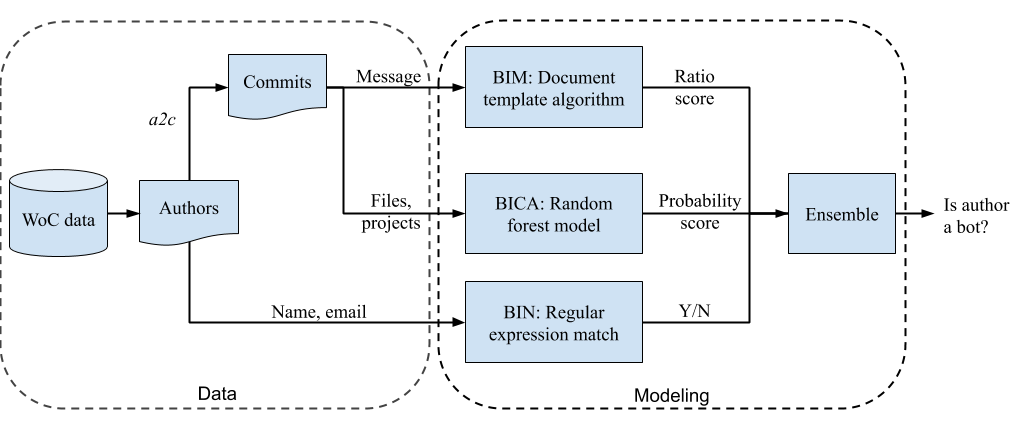}%
\caption{BIMAN workflow: Commit data pertaining to authors is used for message template detection, activity pattern based predictions using a random forest model, and name pattern matching. Scores from each method are used by an ensemble model (another random forest model) that classifies the given author as a bot or not a bot.}
\label{fig:BIMAN}
\vspace{-5pt}
\end{figure*}

In this section, we describe the data used for analysis, present our proposed approach for detecting bots, and describe how we characterized the bots found.

\subsection{Data}\label{ss:data}
The data used for this study was obtained from the World of Code (WoC)~\cite{woc19} dataset. Specifically, version P which was collected between May 15, 2019 and June 5, 2019 based on updates/new repositories identified on May 15, 2019. The data contained information on 73,314,320 unique non-forked Git repositories, 34,424,362 unique author IDs, and 1,685,985,529 commits. The author IDs were represented by a combination of the authors name and email address: \texttt{first-name last-name$<$email-address$>$}. As an example, for an author with first name ``John'', last name ``Doe'', and email address ``myemail@me.com'', the corresponding author ID in the WoC dataset would be: ``\texttt{John Doe<myemail@me.com>}''.

The data is stored in the form of mappings between various Git objects. For our study we used the mappings between the commit authors and commits (\textit{a2c}), commits and filenames (including the file path) changed by that commit (\textit{c2f}), commits and the GitHub projects that commit is associated with (\textit{c2p}), and commits and the contents of the commits comprising the commit timestamp, timezone, and commit message (\textit{c2cc}). 

Our method of extracting information about the authors consisted of the following steps:
\begin{enumerate}
    \item Obtaining a list of all authors from the WoC dataset.
    \item Identifying all commits for the authors using the \textit{a2c} map.
    \item Extracting the list of files modified by a commit, the list of projects the commit is associated with, and the commit content for each of the commits for every author using the \textit{c2f, c2p}, and \textit{c2cc} maps, respectively.
\end{enumerate}

\subsection{Bot Detection}\label{ss:steps}

\textbf{BIMAN}, our proposed technique for detecting bots, comprises three methods, 1) Bot Identification by Name (\textbf{BIN}), 2) Bot Identification by commit Message (\textbf{BIM}), and 3) Bot Identification by Commit Association (\textbf{BICA}), each relying on distinct data attributes.
We discuss these methods separately rather than as a single model because they can be used independently and not all of the required data for each method is available or easily obtainable, and researchers with access to partial data can still use a subset of \textbf{BIMAN}.
An overview of the \textbf{BIMAN} approach is illustrated in Figure~\ref{fig:BIMAN}.

\subsubsection{Identifying bots by name (\textbf{BIN})}\label{sss:bin}
\par
We began devising a possible method for detecting bots by comparing names of known bots. Erlenhov et. al.~\cite{erlenhov2019current} studied 11 bots, which we took as a starting point in our investigation.
However, since the author IDs in our dataset consist of name-email combinations, we had to search through the list of authors for identifying the possible author IDs that could be related to one of these 11 bots. We did not found an entry matching 3 of the 11 bots: ``First-timers'', ``Marbot'', and ``CssRooster'', and found a total of 57 author IDs that could be associated with one of the other bots.
We noticed that 25 (37\%) of these author IDs had the substring ``bot''. We further searched for other known bots (e.g., Travis CI and Jenkins bots) in our dataset and noticed that many of the author IDs we suspected as bots also had the substring ``bot'' in their name or email. 

Based on these observations, regular expressions were used to identify if an author is a bot by checking if the author name or email has the substring ``bot''.
However, to avoid including false positives like ``Abbot'' or ``Botha'', the regular expression searched for ``bot'' preceded and followed by non-alpha characters. 
We further excluded author IDs that had the word ``bot'' only in the domain name of their email addresses (e.g., hr@future-\textit{bot}.ai), since we are not convinced that these are always bots.
Although matching regular expressions does not detect all bots, nor is it able to filter authors trying to disguise themselves as bots, it is a straightforward solution that does not requires any other data, and can be regarded as a good starting point.

\noindent\textbf{Creating the Golden Dataset for \textbf{BIM} and \textbf{BICA}:}
There is no publicly available \textit{golden dataset} of bots in social-coding platforms for training machine learning models.
However, we noticed that the name based bot identification method was very precise, i.e., it had few false positives. Therefore, we used \textbf{BIN} to create a \textit{golden dataset}.
Two of this paper's authors independently analyzed the author IDs and descriptions, and commit and pull request messages, when available, to manually verify the authors identified as bots by \textbf{BIN} and remove the ambiguous cases (less than 1\%) based on consensus. We found a total of 13,150 bot authors via this process.
We also needed to include a set of human authors to complete a training dataset. We randomly selected 13,150 authors, and again manually ensured that no bots were in this list. This was our \textit{golden dataset}, consisting of 26,300 authors, used for training and testing the \textbf{BIM} and \textbf{BICA} methods of \textbf{BIMAN}. 

\noindent\textbf{Comparing the commit activities of humans and bots: }
Our initial assumption was that bots are very active agents and produce a significantly greater number of commits than humans, therefore, we could detect bots by evaluating the number of commits. However, upon investigating the 13,150 bots in the \textit{golden dataset}, we found that assumption to be incorrect. While the maximum number of bot commits was admittedly huge ($2,463,758$), the median number was only $2$, and the first and third quantile values were $1$ and $16$, respectively. In contrast, the median number of human commits was $4$, and the first and third quantile values were $2$ and $17$, respectively. These observations indicated that the number of commits between humans and bots is not significantly different.

We hypothesize that the reason behind why many bots have few commits relates to any of the following reasons: (1) Given that author IDs consist of a name-email combination, slight variations in either appear as different authors, when they are not. For example, a ``dotnet-bot'' 
has three name variations that appear as different authors: \texttt{beep boop}, \texttt{Beep boop}, and \texttt{Beep Boop}, though it has the same email address: \texttt{dotnet-bot@microsoft.com}. We need to employ anti-aliasing methods~\cite{amreen2019alfaa} to address this issue. (2) Bots might have been implemented as an experiment or coursework, and never used afterwards. For example, we found a bot named ``learn.chat.bot'' that most likely belongs in this category. (3) Bots might have been designed for a project, but were never fully adopted.


\subsubsection{Detecting bots by commit messages (\textbf{BIM})}

Characteristics of commit messages can be used to identify an author as a bot.
One approach is to assume that many bots routinely use template messages as the starting point for the commit message.
Consequently, detecting if the commit messages by an author originate from a template can be used to identify such bots. Although humans can also generate commit messages with similar and consistent patterns (e.g., follow a set of software development guidelines), the key assumption \textbf{BIM} follows is that for a large number of commit messages, the variability of messages' content generated by bots is lower than those generated by humans.

\textbf{BIM} utilizes the document template score algorithm presented in Alg. ~\ref{alg:template}.
Given a set of documents (commit messages), the algorithm compares document pairs and uses a similarity measure to group documents.
The \texttt{Similarity} procedure represents a method that computes a ``similarity'' measure that is of interest~\cite{paul2016efficient,helmer2007measuring,buttler2004short,karwath2006relational}, with  the percent identity of the aligned commit messages being used for \textbf{BIM}.
A group represents documents that are suspected to conform to a similar base document, and each group has a single \textit{template document} assigned to it and this is the document always used for comparisons.
A new group is created when a document's similarity with any \textit{template document} does not reach the similarity threshold, $k_b$, and this document is set as the \textit{template document} for that group.
After all documents are compared, a score is calculated based on the ratio of the number of \textit{template documents} and the number of documents: $1 - \frac{\|T\|}{\|D\|}$, where $T$ is the set of \textit{template documents} and $D$ is the set of documents.

In \textbf{BIM}, commit messages were aligned and scored using a combination of global (Needleman-Wunsch~\cite{needleman1970general}) and local (Smith-Waterman~\cite{smith1981identification}) sequence alignment algorithms available via the Python \texttt{alignment}\footnote{\url{https://pypi.org/project/alignment}} library.
The similarity threshold, $k_b$, was set to $40\%$ after testing the accuracy of Alg. ~\ref{alg:template} on the golden data using thresholds of $40$, $50$, $60$, and $70\%$.



\begin{algorithm}
\caption{Document template score}
\label{alg:template}
\begin{algorithmic}[1]
\Statex \textbf{Inputs:} set of documents $D$ and similarity threshold $k_b$
\Statex \textbf{Output:} 1 - ratio of number of templates to documents
    \State $T \leftarrow \emptyset$ \Comment{set of template documents}
    \State $\bm{G} \leftarrow \{\emptyset\}$ \Comment{template groups, $\bm{G}_i$ is associated to template $i$}
    \For {$d \in D$}
        \For {$t \in T$ \textbf{and} $d \notin \bm{G}$}
            \If {\textsc{Similarity}($d$, $t$) $> k_b$}
                \State Add $d$ to $\bm{G}_{t}$
            \EndIf
        \EndFor
        \If {$d \notin \bm{G}$}
            \State Add $d$ to $T$
            \State Add $d$ to $\bm{G}_{d}$
        \EndIf
    \EndFor
    \State \textbf{return} $1 - \frac{\|T\|}{\|D\|}$
\end{algorithmic}
\end{algorithm}
\vspace{-5pt}


\subsubsection{Detecting bots by files changed and projects associated with commits (\textbf{BICA})}\label{sss:bica}

We calculated $20$ metrics using the files changed by each commit, the projects that commit is associated with, and the timestamp and timezone of the commits, based on our initial assumptions about how bots and humans might be different, and empirical validation of the assumption by observing the differences in distribution of those variables for bots and humans. 

For predicting whether an author is a bot using the numerical features, we tested several modeling approaches: linear and logistic regression, generalized additive models, support vector machines, and random forest. The random forest model performed better than the other approaches, so we decided to use that approach. We used the random forest implementation available in the ``randomForest'' package in R, with these $20$ variables as predictors, to predict if the author of those commits is a bot. After iteratively selecting and removing predictors based on their importance in the model, and measuring the AUC-ROC every time, we found that a model with only $6$ predictors was the best model. The list of predictors is given in Table~\ref{t:var}, along with the description of each variable. We found that the timestamp of a commit and any time related measure (e.g., how long a bot has been active and at what times of the day it makes commits) are not important predictors.

In order to tune the random forest model, we used the ``train'' function from the \textit{caret} package in R for performing a grid search (using a $10$ fold cross-validation) on the training data to find the best values of the model parameters that resulted with the highest accuracy: ``ntree" (number of trees to grow) and ``mtry" (number of variables randomly sampled as candidates at each split). The optimum values for ``ntree'' and ``mtry'' were $100$ and $2$, respectively.

\begin{table}[!htb]
\caption{Predictors used in the random forest model}
\label{t:var}
\resizebox{0.8\linewidth}{!}{%
\begin{tabular}{@{}p{2cm}p{4.75cm}@{}}
\toprule
\multicolumn{1}{c}{\textbf{Variable Name}} & \multicolumn{1}{c}{\textbf{Variable Description}} \\ \midrule
Tot.FilesChanged & Number of files changed by author across commits (includes duplicates) \\\hline
Uniq.File.Exten & Number of unique file extensions in all the author's commits \\\hline
Std.File.pCommit & Std. dev. of number of files per commit \\\hline
Avg.File.pCommit & Mean number of files per commit \\\hline
Tot.uniq.Projects & Number of unique projects commits have been associated with \\\hline
Median.Project. pCommit & Median number of projects the commits have been associated with (includes duplicates); We took the median value, because the distribution of projects per commit was very skewed, and the mean was heavily influenced by the maximum value.  \\ \bottomrule
\end{tabular}%
}
\vspace{-5pt}
\end{table}

\subsubsection{Ensemble model:}\label{sss:ens} \par
Based on the fact that \textbf{BIN}, \textbf{BIM}, and \textbf{BICA} methods consider different aspects of the authors and commits, we decided to use an ensemble model, implemented as another random forest model.
The ensemble model in \textbf{BIMAN} utilizes the outputs of the three methods as predictors to make a final judgement as to whether an author is a bot or not. The output from \textbf{BIN} is a binary value ($1\rightarrow$ bot, $0\rightarrow$ human), stating if the author ID matches the regular expressions we checked against; the output from \textbf{BIM} is a score, with higher values corresponding to a higher probability of the author being a bot; and the output from \textbf{BICA} is the probability that an author is a bot. 

\noindent\textbf{Creating the Training Dataset for the ensemble model:}
Recall the \textit{golden dataset} was generated using the \textbf{BIN} method, so we did not used it for training the ensemble model. Instead, we created a new training dataset partly consisting of 67 bots from which 57 author IDs were associated with 8 bots described in~\cite{erlenhov2019current} (as mentioned in Section~\ref{sss:bin}) and 10 author IDs associated with 3 other known bots that were not in the \textit{golden dataset}: \textit{Scala Steward}, \textit{codacy-badger}, and  \textit{fossabot}. Also, 67 human authors were included via random selection and manual validation. The final training data for the ensemble model had only 134 observations, however, given that we had 3 predictors, we were reasonably satisfied with it.

\subsection{Bot Characterization}\label{ss:charcterize}
Instead of trying to design a taxonomy of bots, as was done in previous studies~\cite{erlenhov2019current,lebeuf2018taxonomy}, 
we aim to classify bots by their activity patterns, specifically, by the files modified and the timestamps of commits, due to the limited information available of the bots. This characterization was applied to the 13,150 bots in the \textit{golden dataset}. We looked at how the number of commits made by a bot are distributed over the 24 hours of a day, which gives an indication of what type of work a bot performs.


\begin{table*}[!htb]
\caption{Performance of the models in detecting 8 known bots from~\cite{erlenhov2019current}  and 3 other known bots outside the \textit{golden dataset}}
\label{t:bots}
\resizebox{0.85\textwidth}{!}{%
\begin{tabular}{@{}lp{3cm}p{3cm}p{3cm}p{3cm}p{3cm}@{}}
\toprule
\small{\textbf{Bot}} & \small{\textbf{No. of author IDs associated with the bot}} & \small{\textbf{No. of IDs identified as bot by BIN}} & \small{\textbf{No. of IDs identified as bot by BICA}} & \small{\textbf{No. of IDs identified as bot by BIM}} &  \small{\textbf{No. of IDs identified as bot by BIMAN}} \\ 
\midrule
Dependabot & \large{4} & \large{4} & \large{4} & \large{2} & \large{4} \\
Greenkeeper & \large{15} & \large{10} & \large{13} & \large{11} & \large{13} \\
Spotbot & \large{1} & \large{0} & \large{1} & \large{1} & \large{1} \\
Imgbot & \large{5} & \large{1} & \large{4} & \large{3} & \large{4} \\
Deploybot & \large{29} & \large{9} & \large{20} & \large{17} & \large{23} \\
Repairnator & \large{1} & \large{0} & \large{0} & \large{1} & \large{1} \\
Mary-Poppins & \large{1} & \large{0} & \large{1} & \large{1} & \large{1} \\
Typot & \large{1} & \large{1} & \large{0} & \large{1} & \large{1} \\ \hline
Scala Steward & \large{6} & \large{0} & \large{6} & \large{6} & \large{6} \\
codacy-badger & \large{2} & \large{0} & \large{2} & \large{2} & \large{2} \\
fossabot & \large{2} & \large{0} & \large{2} & \large{2} & \large{2} \\ \hline
\textbf{Total} & \textbf{67} & \textbf{25 (37\%)} & \textbf{53 (79\%)} & \textbf{47 (70\%)} & \textbf{58 (87\%)}\\
\bottomrule
\end{tabular}%
}
\vspace{-5pt}
\end{table*}

\subsubsection{Characterization of bots by activity hours:}
By observing the distribution of commits created by a bot over the 24 hours of a day using time-of-day histograms for a randomly selected sample of 50 bots identified by \textbf{BIMAN}, and employing a qualitative analysis technique similar to card sorting, we identified three distinct patterns: 1) Bots were active almost uniformly over the 24 hours of a day, or, in some cases, they had no activity for a few contiguous hours and almost uniform activity during the rest of the day; 2) Bots' activity patterns resembled the typical activity patterns of humans very closely, i.e., they were more active during ``working hours'', with a few contiguous peak activity hours, and they had limited activity over the rest of the day; 3) Bots were active only for a few specific hours and had almost no activity during the rest of the day. We named the three types of bot patterns, respectively, as: ``Continuous Activity Bots'', ``Synchronous Activity Bots'' (their activity seems to be synchronized with typical human activity), and ``Spike Activity Bots''. There were some bots with activity patterns that did follow any of the three patterns described, we classified these bots as ``Other Bots''. 

We applied this characterization only on the ``active'' bots, because bots with very few commits would almost always follow the ``Spike Activity Bots'' or ``Other Bots'' pattern. Moreover, we were more interested in the ``active'' bots because they have a greater influence over the projects and ecosystems they are active in, as they have more code contribution than the rest. We designated the bots with more than 1,000 commits as ``active'' bots, and we found 454 (3\%) bots in the \textit{golden dataset} matching that criteria. For the characterization process, we had three of this paper's authors independently classify the bots based on their commits' distribution over the hours of day and a fourth author compile the results into a final classification, taking into account the rating stated by the others and using her own judgement for the ambiguous cases.

\subsubsection{Identifying files modified by bots:}
We also investigated which programming languages bots worked with and what types of files they modified. We extracted the file extensions from the files modified by each commit from the bots in the \textit{golden dataset}, and used the \textit{linguist}~\footnote{\url{https://github.com/github/linguist}} tool to obtain an estimated language classification based on a common open-source model. This information was used to infer which programming languages a bot worked with.

\section{Results}\label{s:result}

\subsection{Qualitative Validation of \textbf{BIMAN}}\label{ss:qval}


Before going into the detailed performance evaluation of \textbf{BIMAN}, we wanted to test how it performs in detecting a few known bots. As mentioned in Section~\ref{sss:bin}, we obtained a set of 57 author IDs associated with 8 of the bots described in~\cite{erlenhov2019current}. In addition, we examined 10 author IDs associated with 3 well-known bots,  \textit{Scala Steward}, \textit{codacy-badger}, and  \textit{fossabot}, that were not in the \textit{golden dataset}. 
The performance of bot detection of \textbf{BIMAN} and each of its constituent methods is shown in Table~\ref{t:bots}.

We found that \textbf{BIMAN} identified 58 (87\%) out of 67 author IDs as bots, and 6 out of 9 other IDs could be identified as not actually being a bot via manual investigation, they were either spoofing the name or simply using the same name. The 3 other IDs, 2 of which were associated with ``Deploybot''~\footnote{deploybot-lm <45803032+deploybot-lm@users.noreply.github.com>}~\footnote{DeployBot <deploybot@imqs.co.za> }, and the other with ``Imgbot''~\footnote{imgbot<imgbothelp@gmail.com>}, had 1 commit each, making any decision about them being bots difficult to make even via manual investigation. 

\subsection{Performance Evaluation of \textbf{BIMAN}}

In this section, we discuss the performance of \textbf{BIMAN}, our proposed approach for bot detection. As mentioned in Section~\ref{ss:steps}, \textbf{BIMAN} is comprised of three independent methods, each looking at a different aspect of the commit authors and the commits, and an ensemble model that combines the results from the three methods for estimating the final prediction. We decided to evaluate the performance of each method and discuss what we learned with each one in detecting bots that commit code.

\subsubsection{Performance of \textbf{BIN}:} 
We did not use the \textit{golden dataset} to validate the accuracy of \textbf{BIN} because this method was used to construct that dataset (see Section~\ref{sss:bin}). However, during creation of the \textit{golden dataset}, \textbf{BIN} obtained a precision close to 99\%, which indicates that any author considered to be a bot using this method has a very high probability of being a bot. In general, humans do not try to disguise themselves as bots. The recall measure is not high, because \textbf{BIN} missed a lot of cases where the bots do not explicitly have the substring ``bot'' in their name. As mentioned in Section~\ref{ss:qval}, we observed an estimated recall of 37\% on the set of 67 bot IDs we manually investigated.

\subsubsection{Performance of \textbf{BIM}:}

\begin{figure*}[t!]
        \centering
        \begin{minipage}[c]{0.3\textwidth}
            \centering
            \noindent\includegraphics[width=\textwidth]{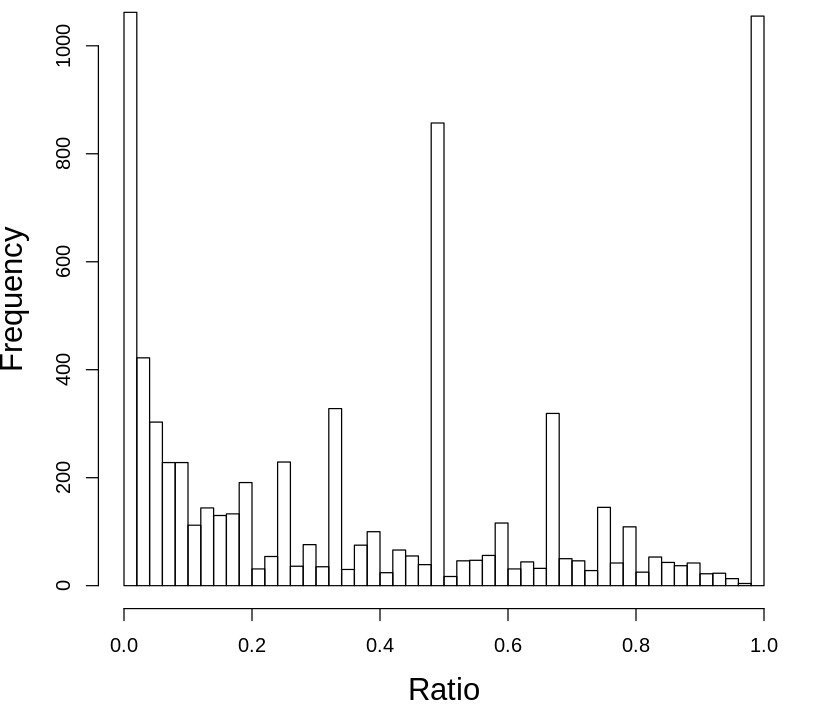}
            (a)
        \end{minipage}
        \begin{minipage}[c]{0.3\textwidth}
            \centering
            \noindent\includegraphics[width=\textwidth]{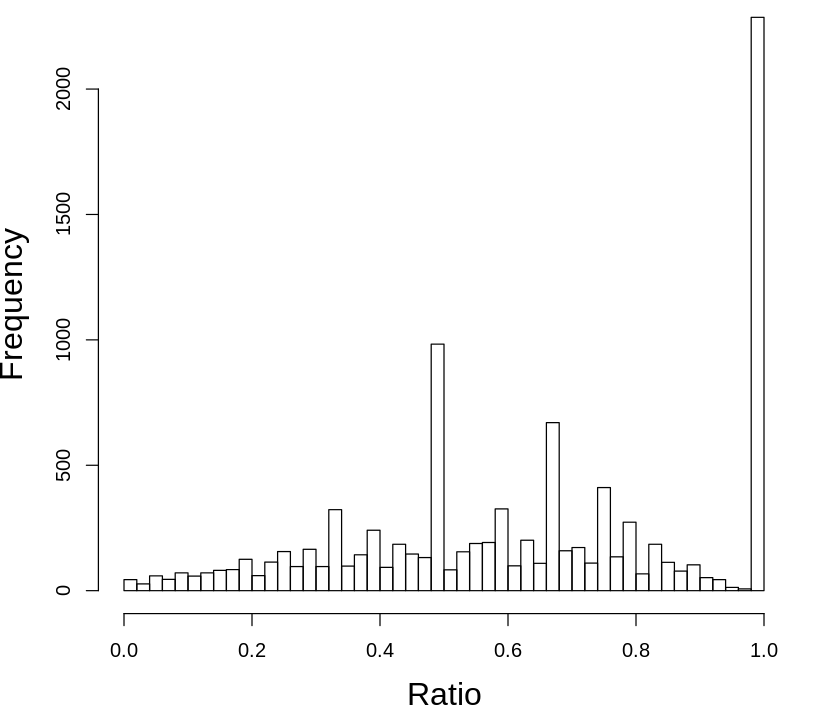}
            (b)
        \end{minipage}
        \begin{minipage}[c]{0.25\textwidth}
            \centering
            \noindent\includegraphics[width=\textwidth]{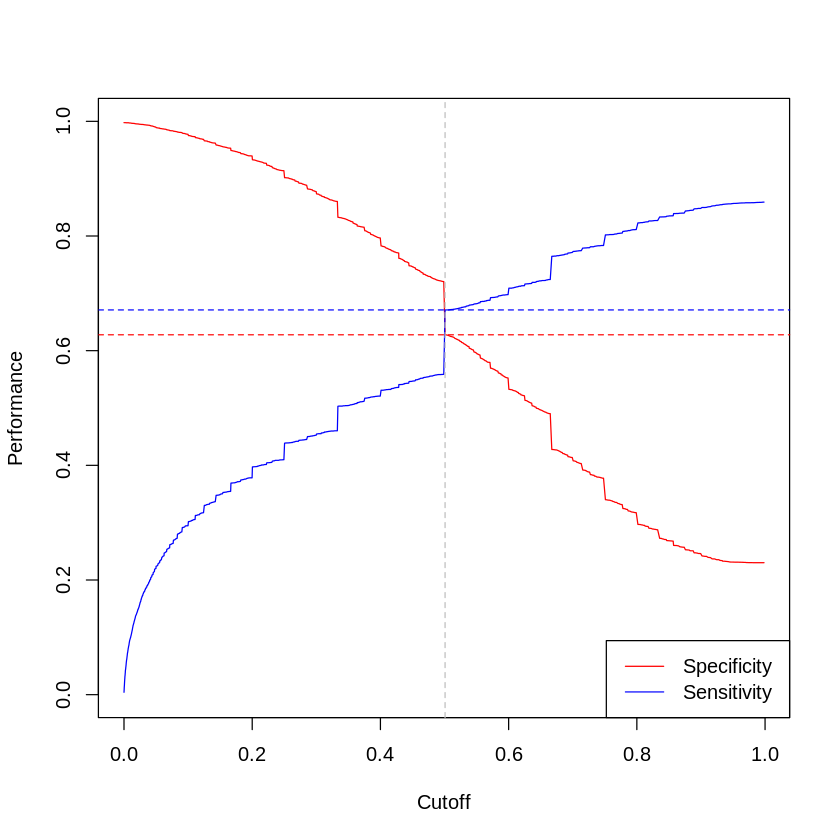}
            (c)
        \end{minipage}
        \caption{(a) Ratio of number of detected templates and the number of commit messages for the 13,150 bots in the \textit{golden dataset}; (b) Ratio of number of detected templates and the number of commit messages for the 13,150 humans (non-bots) in the \textit{golden dataset}; (c) Plot of sensitivity, specificity, and cutoff (threshold), when predicting if an author is a bot using the ratio of number of detected templates to the number of commit messages.}
        \label{fig:ratio}
\end{figure*}



Our proposed method for detecting whether an author is a bot by applying the document template score algorithm, Alg.~\ref{alg:template}, solely relied on the commit messages. Figures~\ref{fig:ratio}a-b present the ratio of the number of probable templates to the number of commit messages for the bots and humans in the \textit{golden dataset}. Note that bots tend to have a lower ratio than humans. The reason for both plots having a high template ratio is that if an author has a single commit message, the ratio is bound to be 1. Over 25\% of the bots in the \textit{golden dataset} have only 1 commit. 

We wanted to find an optimal threshold for the template ratio, so that the authors, for whom the ratio is lower than the threshold, can be regarded as bots (the output from the \textbf{BIM} method is $(1-$ratio), so a lower value is more likely to be human and vice versa). This information would be useful for researchers who might only use this technique to detect whether a given author is a bot, and this also helps us calculate the performance of this method. The optimal threshold was found using the ``closest.topleft'' optimality criterion (calculated as: $min((1 - sensitivities)^2 + (1- specificities)^2)$) using the \textit{pROC} package in R. The AUC-ROC value using the ratio values as predicted probabilities was 0.70, and the optimal values for the threshold, sensitivity, and specificity were found to be 0.51, 0.67, and 0.63, respectively. 
We plotted the sensitivity and the specificity measures in Figure~\ref{fig:ratio}c, and highlighted the optimal threshold, sensitivity, and specificity values for that threshold.


\noindent\textbf{True Positive:} The cases where this model could correctly identify bots were cases where the bots actually used templates or repeated the same commit message, e.g., a bot named ``Autobuild bot on Travis-CI'' used the same commit message ``\texttt{update html,}'' for all of the 739 commits it made, and a bot named ``Common Workflow Language project bot'' created 1,373 commits that used the form: ``\texttt{\$USER-CODE: \$SOFTWARE configuration files updated. Change performed by \$NAME}''. \textbf{BIM} could identify these messages as coming from the same template message and classify these authors as bots.

\noindent\textbf{False Negative:} The cases where this model could not correctly identify bots were mostly cases where the bots reviewed code added by humans and created a commit message that added a few words with the commit message written by a human, e.g., a bot named ``Auto Roll Bot'' created commit messages in the form of: ``\texttt{\$COMMIT-SEQUENCE-NUMBER: \$LONG-HUMAN-COMMIT-TEXT\\ \$PATTERN}'', 
with one specific example being ``\texttt{3602: Fix errors
in the Newspeak Mac installer genrators. Fix a slip in
platforms/Cross/vm/\\
sqCogStackAlignment.h for the ARM's getsp. Eliminate
non-spur and stack VMs from the ARM builds (it builds
veerry slowly) Include 64-bit and Mac Pharo VMs in
archives and uploads.-------------------------------}'',
with the length of ``\$LONG-HUMAN-COMMIT-TEXT'' typically ranging between 20 and 50 words. \textbf{BIM} failed to identify this template and misclassified this author as a human.

\noindent\textbf{True Negative:} The human authors correctly identified had some variation in the text, with the usual descriptions of change. Some examples are: ``\texttt{Added a count down controller}'' and ``\texttt{Enabling multiport deployments. By mapping ports a little bit more specific we get all the servers listed in the server browser}''. 

\noindent\textbf{False Positive:} In contrast, humans who were misclassified as bots usually had short commit messages that were not descriptive, and they reused the same commit message multiple times. Example of typical messages are: ``\texttt{Initial Commit}'', ``\texttt{Added File by Upload}'', and ``\texttt{Updated \$FILE}''. 

Our observations support that \textbf{BIM} is useful in detecting ``typical'' bots that modify small parts of a message in every commit, and ``typical'' humans who write descriptive commit messages. However, we can also conclude that it is very hard to identify if an author is a bot using just one signal.

\subsubsection{Performance of \textbf{BICA}:}

As mentioned in Section~\ref{sss:bica}, the \textbf{BICA} approach uses a random forest model with the measures listed in Table~\ref{t:var} as predictors for identifying bots. We used the \textit{golden dataset} generated using the \textbf{BIN} method (Section~\ref{sss:bin}) for training the model and testing its performance. 70\% of the data, selected randomly, was used for training the model and the rest 30\% was used for testing it, and the procedure was repeated 100 times with different random seeds.

\begin{figure}[!t]
\centering
\includegraphics[width=0.65\linewidth]{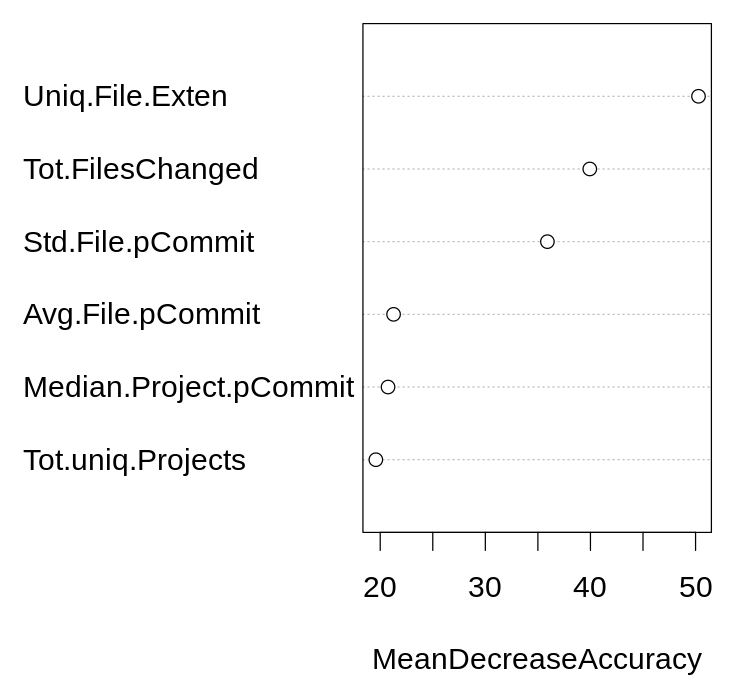}%
\caption{Variable importance plot for the random forest model used to identify bots}
\label{fig:rf-vimp}
\end{figure}

The model showed good performance, with an AUC-ROC value of $0.89$. The variable importance plot (Figure~\ref{fig:rf-vimp}) indicates that the total number of unique file extensions and the total number of files changed in all the commits made by an author are the most important variables.

\begin{figure}[!t]
\centering
\includegraphics[width=\linewidth]{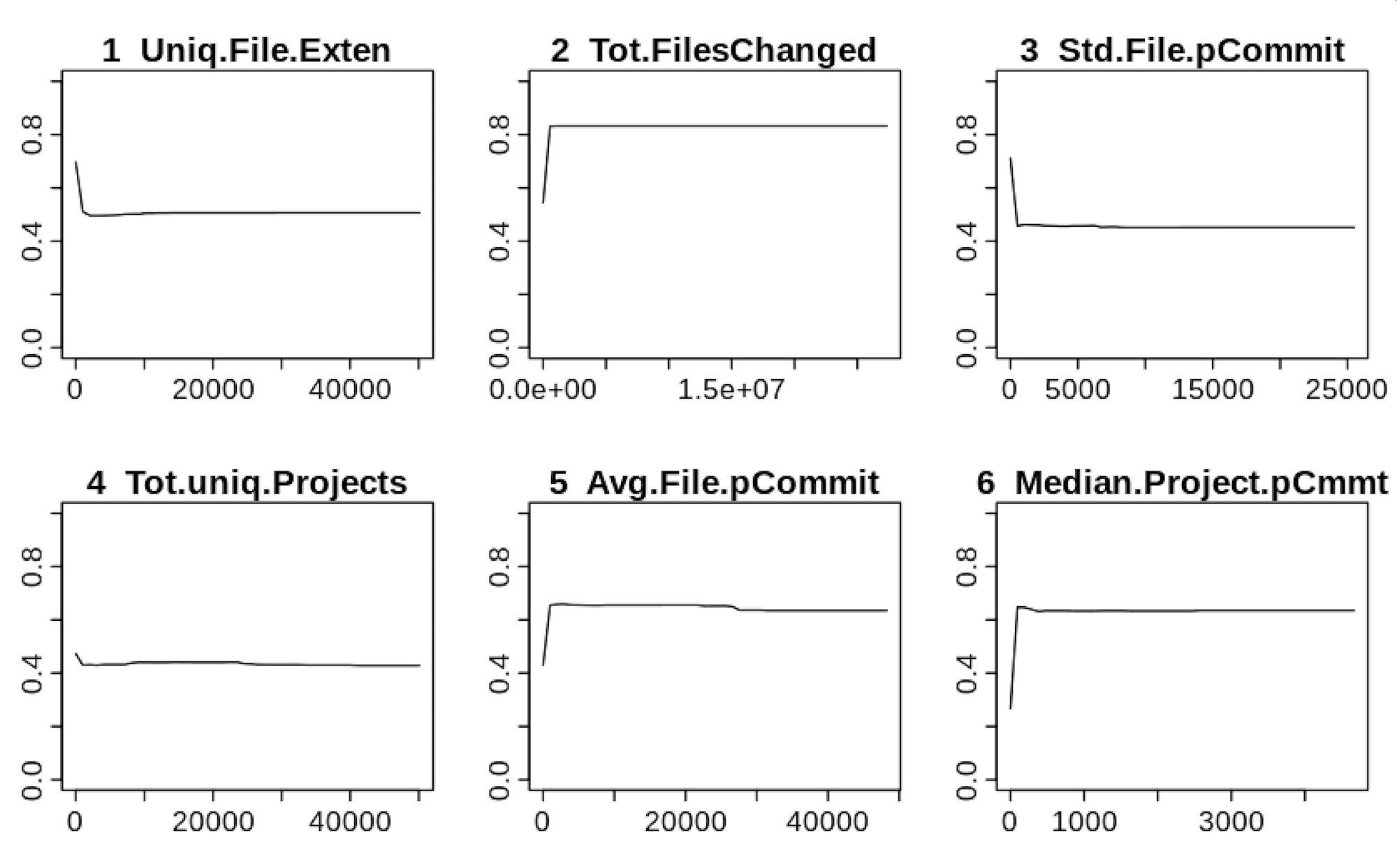}%
\caption{Partial dependence plot for all the predictors in the random forest model }
\label{fig:rf-pd}
\end{figure}

To understand what each of the predictors tell us about how the behavior of the bots differ from that of humans, we looked at their partial dependence plots, see Figure~\ref{fig:rf-pd}. The greater values in the vertical axis of each plot correspond to a higher probability of an author being a bot, and the values in the horizontal axis are the possible predictors' values. These plots illustrate an empirical understanding about how the behavior of the bots is different from humans. We notice that bots tend to have fewer number of unique file extensions and their commits are associated with fewer number of different projects, i.e., they tend to operate in one ecosystem. However, their commits tend to be associated with a greater number of projects per commit, the projects they commit to are more popular. 
Bots typically make larger commits, as we notice that they tend to have more files per commit on average and a greater number of total files changed. They are also more consistent in terms of commit size because the variation in the number of files per commit is lower. These observations fall in line with our idea of typical bots, which keep updating a consistent set of files and typically partake in popular projects.

\subsubsection{Performance of the ensemble model:}
We combined the results of \textbf{BIN}, \textbf{BIM}, and \textbf{BICA} using an ensemble model, implemented as a random forest model. The dataset used for training and testing the performance of this model had only 134 observations, because of reasons described in Section~\ref{sss:ens}. We used 80\% of the data for training, and 20\% for testing, and repeated the process 100 times with different random seeds. The performance of this model had variation because of the small size of the training data. The value of the AUC-ROC measure varied between 0.89 and 0.95, with a median of 0.90. 

\hypobox{
To address RQ1, we devised \textbf{BIMAN}, a systematic approach for detecting bots using information about their names, commit messages, files modified by the commit, and projects associated with the commits. 
}

\subsection{Estimating the Number of Commit Bots}\label{ss:estim}
While we can easily obtain the number and activity of author IDs that contain the substring ``bot'', it is much  
more difficult to determine the total number of author IDs that, from their string representation, can not be inferred to be bots. 
Yet, even a rough gauge on the prevalence of bots and code committed by bots would be helpful to have a handle on 
the fraction of code commit activity that is automated.
To do that we randomly selected a sample of 10,000 authors IDs outside of our datasets used so far (none had the substring ``bot'' in their names). \textbf{BIMAN} predicted 1,167 authors IDs to be bots, and we randomly sampled 100 authors IDs among those. 
A subjective assessment conducted by two authors of this paper discovered at least 9 
of these author IDs likely to produce mostly automated commits. From this, we can obtain a rough estimate that approximately $11.67\% \times 9\% \approx 1\%$ of all authors IDs who commit code are bots. Therefore, from the total population of approximately 40 million 
authors in open-source Git commits, approximately 400,000 authors are bots. The 9 author IDs that we 
identified as bots were found to have created between 10 and 1,500 times more commits than the remaining author IDs. Such high discrepancy 
strengthens our concerns described in Section~\ref{s:motiv} that the empirical analyses relying on measures of developer 
productivity can be strongly biased even if the actual bot population represents a modest $1$\% of all developer author IDs.

\subsection{Shared Dataset of Bot Commits}
We have compiled the information about the commits made by $461$ bots detected using \textbf{BIMAN}, each of whom have created more than 1,000 commits, into a single dataset and made it available for researchers interested in conducting studies on such data, which includes information about $13,762,430$ commits made by these bots. We decided to focus on the more active bots since these bots would have a much greater effect on any estimate of developer productivity, team size, etc. and they are the ones that should be accounted for during any data cleaning process.

The data is stored in a delimited text file (semicolon as the separator) with the following format in each line: ``author\_id''; ``commit-sha''; ``time-of-the-commit''; ``timezone''; ``files-modified-by-the-commit''; ``projects-the-commit-is-associated-with''; ``commit-message''. In the case of having multiple projects and files for a given commit, they are separated by ','. 
The data is available at \textit{Zenodo}, through the link provided in Section \ref{s:intro}. Additional data about other authors, along with the likelihood of each being a bot will be provided upon requests.

\subsection{Bot Characterization}


\subsubsection{Characterizing the bots based on their activity during the day: }

\begin{figure*}[t!]
        \centering
        \begin{minipage}[c]{0.22\textwidth}
            \centering
            \noindent\includegraphics[width=\textwidth]{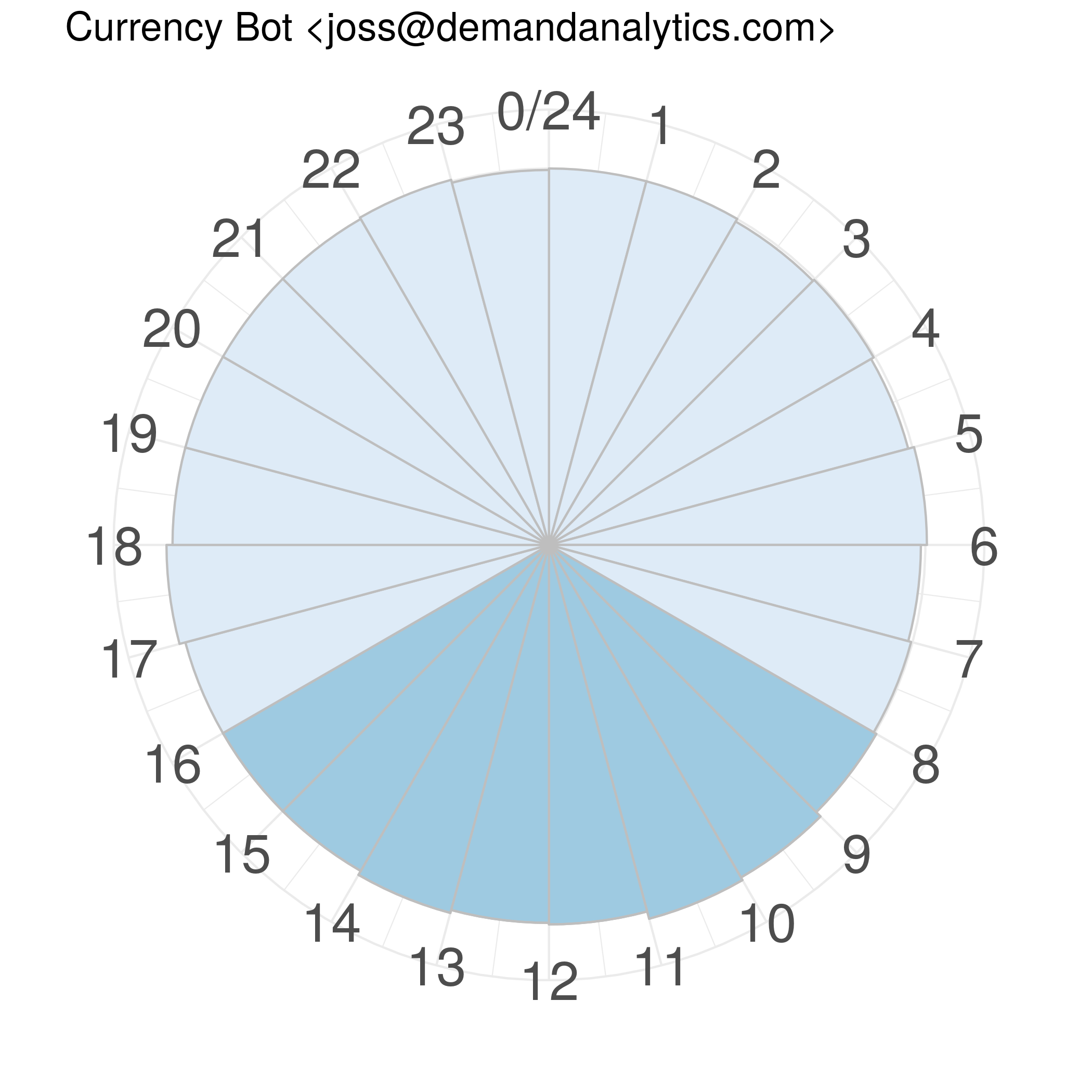}
            (a)
        \end{minipage}
        \begin{minipage}[c]{0.22\textwidth}
            \centering
            \noindent\includegraphics[width=\textwidth]{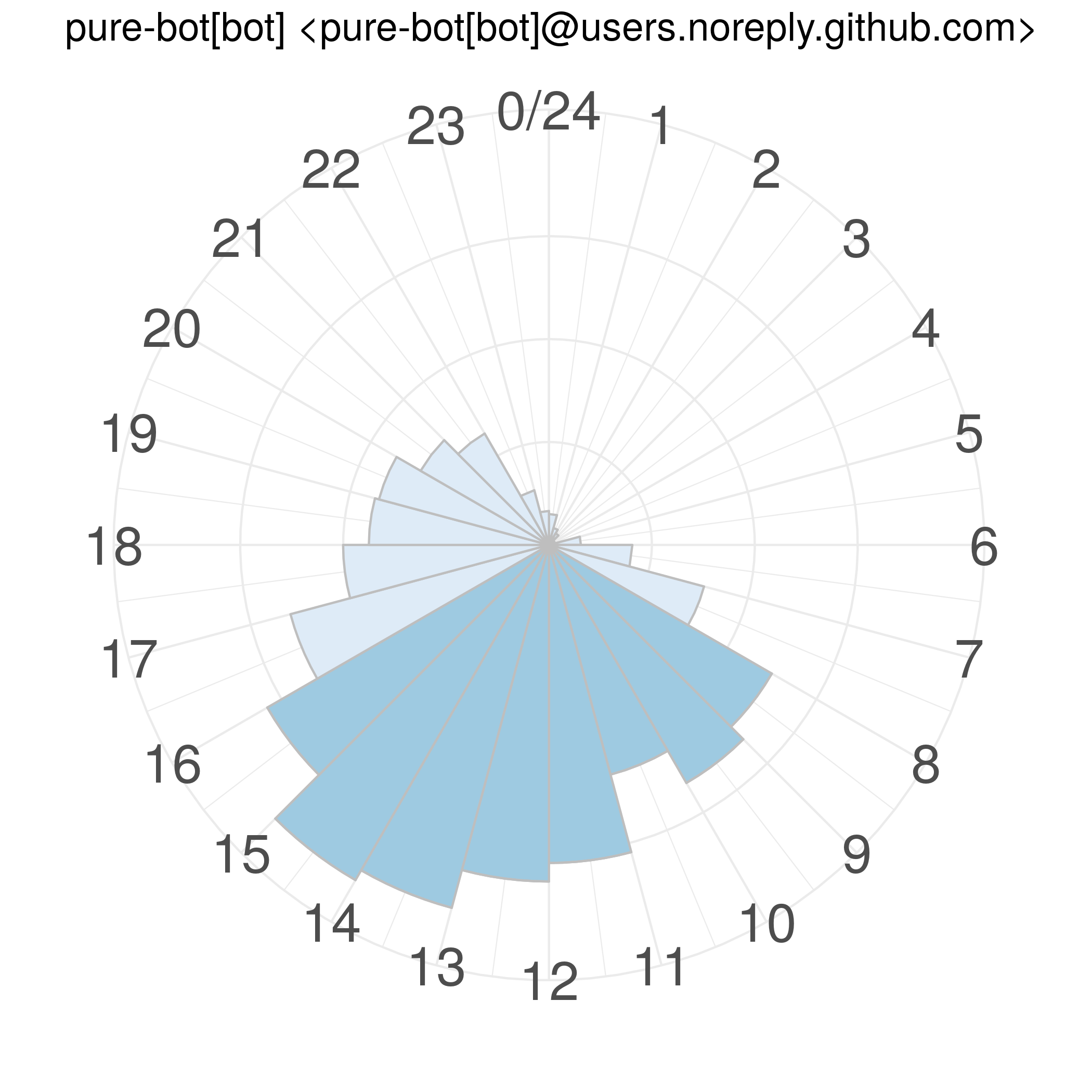}
            (b)
        \end{minipage}
        \begin{minipage}[c]{0.22\textwidth}
            \centering
            \noindent\includegraphics[width=\textwidth]{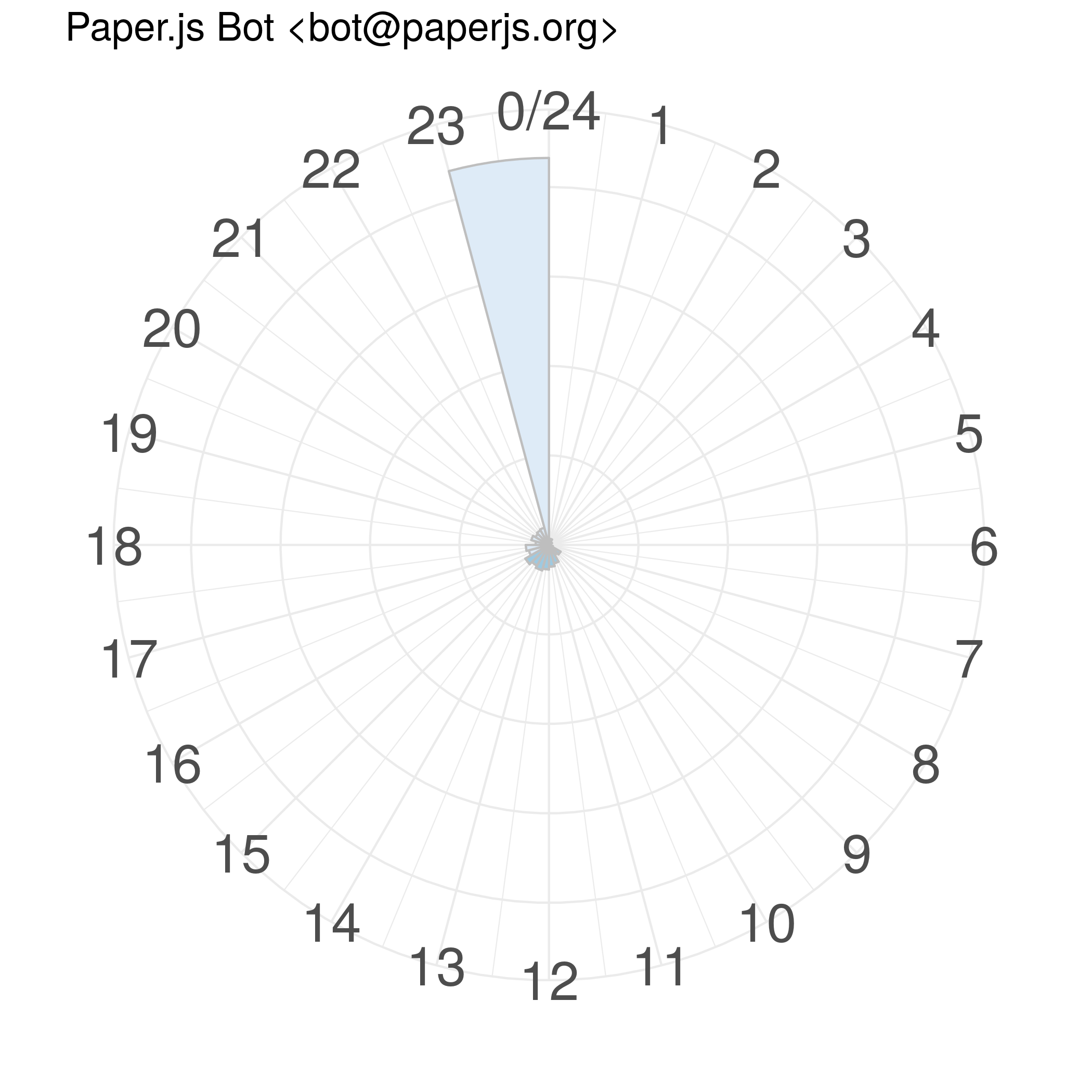}
            (c)
        \end{minipage}
        \begin{minipage}[c]{0.22\textwidth}
            \centering
            \noindent\includegraphics[width=\textwidth]{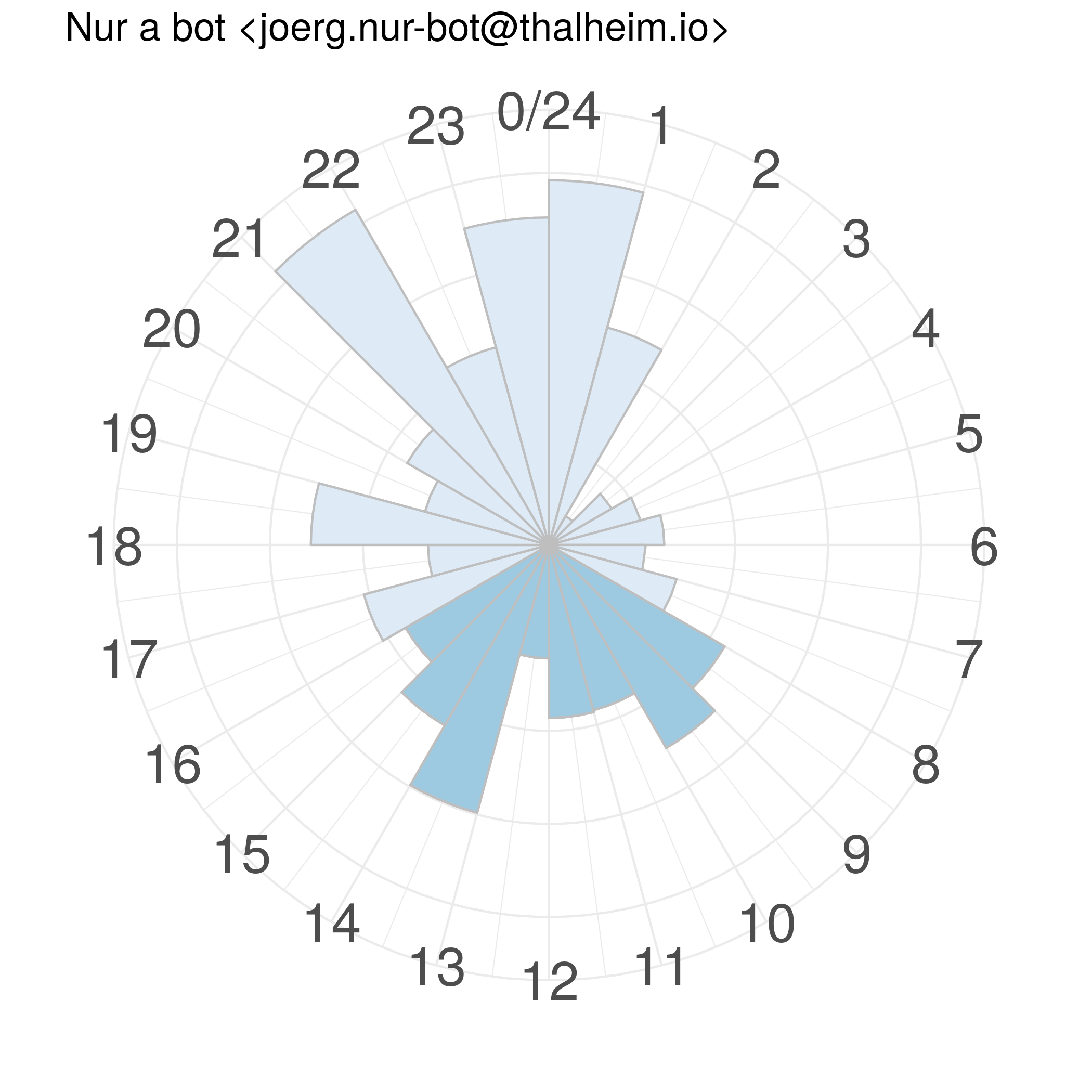}
            (d)
        \end{minipage}
        \caption{Classes of bots identified via activity patterns: (a) The Currency Bot - A Continuous Activity Bot, (b) The Pure Bot - A Sync Activity Bot, (c) The Paper.js Bot - A Spike Activity Bot, (d) Nur a Bot - An Other Bot }
        \label{fig:results}
\end{figure*}





As mentioned in Section~\ref{ss:charcterize}, we classified the bots into 3 particular categories, and an ``others'' category for bots that do not fall in any of the previous ones. After going through the categorization process for the 454 bots with more than 1,000 commits, as mentioned in Section~\ref{ss:charcterize}, we found 100 (22\%)  ``continuous activity pattern bots'',  162 (36\%) ``synchronous activity pattern bots'', 128 (28\%) ``spike activity pattern bots'', and 64 (14\%) bots classified as ``others''.  

Here we give one typical example from each category, and describe what type of work they do. This would help in understanding why these bots have the observed activity pattern, and would enable inferring what type of the work a bot does if it falls in one of the observed categories. We also show how their commit activity over the 24 hours of a day are distributed using radial activity plots, where we show what is the relative amount of activity of the bot in a given hour over its lifetime. The hours between 8 a.m. and 4 p.m. are highlighted in each plot, since these hours are known to be the typical working hours. The plots in Figure~\ref{fig:results} are all examples of radial activity plots.

\noindent\textbf{1. Continuous Activity Bots: }
A continuous activity bot shows almost uniform activity over the 24 hours of a day. A representative example of this class is the ``Currency bot'',~\footnote{\url{https://github.com/currencybot}} which collects up-to-date exchange rate data every hour and distributes it for free. The radial activity plot for this bot is shown in Figure~\ref{fig:results}a. This bot has a very uniform distribution of activity because it is active throughout the day. Other bots in this category also perform tasks that require them to be similarly active throughout the day.

\noindent\textbf{2. Sync Activity Bots: } 
These bots typically work in response to the activity of a human, which means their activity during the day closely resembles that of a human. A typical example of the Sync bot class is the ``Pure bot''~\footnote{\url{https://github.com/syndesisio/pure-bot}}, which enables automated pull request workflows, reacting to input from web-hooks, and performing actions as configured. The radial activity pattern for this bot, as shown in Figure~\ref{fig:results}b, is similar to the typical activity pattern of humans because it reacts in response to humans creating pull requests. 

\noindent\textbf{3. Spike Activity Bots: } 
These bots perform actions at a fixed time of the day or at regular intervals. The type of work they perform primarily falls into two categories, backup data and update websites. A prime example of such a bot is the ``Paper.js Bot''~\footnote{\url{https://github.com/paperjs-bot}}, which automatically regenerates the Paper.js website once per day, and its radial activity plot is shown in Figure~\ref{fig:results}c.

\noindent\textbf{4. Other Bots: } There are a number of bots whose activity pattern do not match any of the categories previously described. A particular example is the ``Nur a Bot''~\footnote{\url{https://github.com/nur-bot}}, which is responsible for updating the NUR repository based on community updates and NIX repository updates. Nix is a powerful package manager for Linux and other Unix systems that makes package management reliable and reproducible, and NUR is a community repository where anyone can submit software to be added in. Seeing as this bot does two different types of work, its activity does not follows a regular pattern, as can be observed by its radial activity plot in Figure~\ref{fig:results}d. 

Our effort of categorizing bots code sheds light on which specific activities they do, thus it covers the ``Initiation'' facet described in~\cite{erlenhov2019current}. Broadly speaking, we can infer that the bots that show continuous activity or spike-like activity are the ``active'' bots, which are activated by the system running it. In contrast, the ``Synchronous Activity Bots'' are ``reactive'', i.e., they work in response to the activity of a human or another bot. We can not infer anything specific about the ``Other'' bots with the information we have.

\subsubsection{What types of files do bots modify?}

\begin{figure}[!t]
\centering
\includegraphics[width=0.58\linewidth]{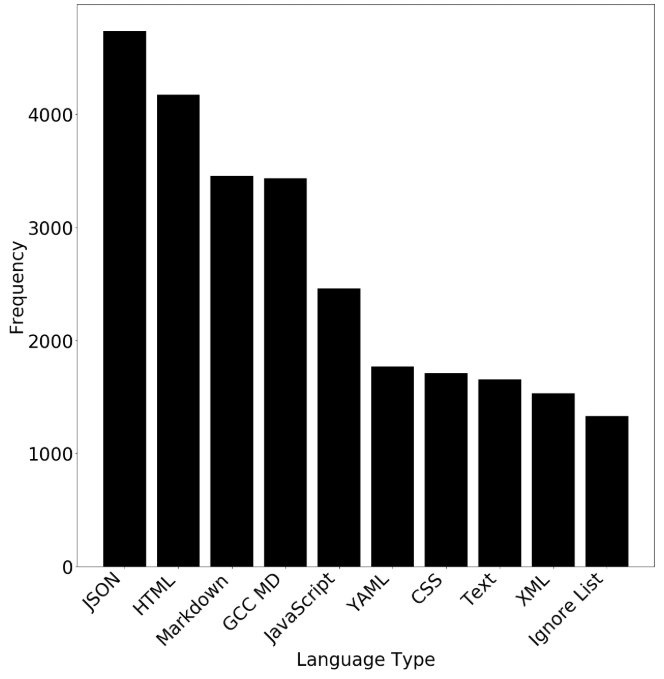}%
\caption{The types of files most frequently modified by bots}
\label{fig:file}
\end{figure}

\begin{figure}[!t]
\centering
\includegraphics[width=0.58\linewidth]{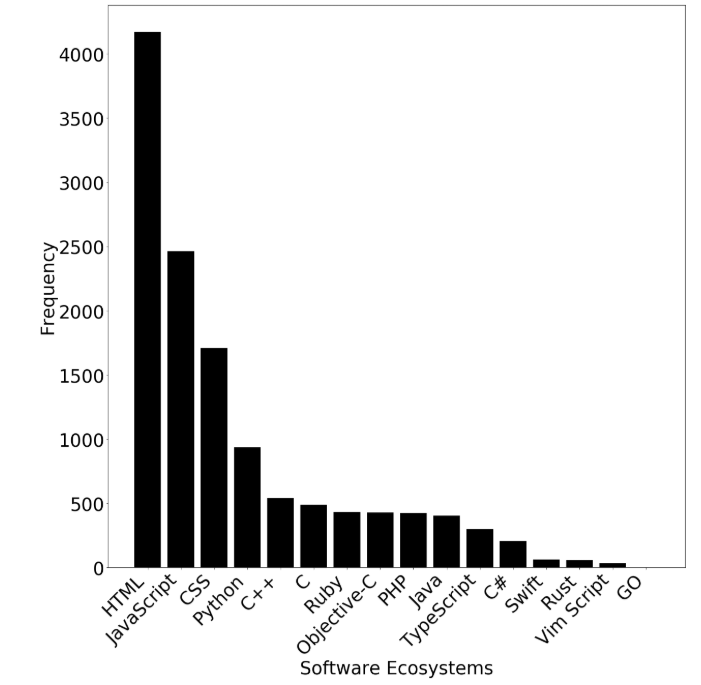}%
\caption{The programming languages bots contribute to}
\label{fig:eco}
\end{figure}

Understanding the types of files these bots work with can give us more insight into the type of work they perform and which programming languages they work with. 
Following the characterization steps as described in Section~\ref{ss:charcterize}, we discovered the types of files modified by the 13,150 bots in the \textit{golden dataset}. The type of files modified most frequently by bots are shown in Figure~\ref{fig:file}. The frequency values in the vertical axis of the plots represent the number of bots that have modified a specific type of file. We notice that bots mostly work with configuration and documentation related files, along with HTML and JavaScript files. 

We also tried to investigate which programming languages bots most frequently work with, so we took the list of languages examined by Wessel et. al.~\cite{wessel2018power} and measured how many of the 13,150 bots have contributed to one of them. The distribution of bots in different languages is shown in Figure~\ref{fig:eco}, with the number of bots working with a particular language shown in the vertical axis. We notice that, similar to what we observed earlier, HTML and JavaScript are the languages bots are most active in, which corroborates the findings of~\cite{wessel2018power}.

\hypobox{Regarding RQ2, we identified four different types of bots based on their commit activity during the day and found that bots primarily work with configuration and documentation files, with HTML and JavaScript being the most common programming languages they contribute to.}

\section{Limitations and Future Work}\label{s:limit}
Our approach of detecting bots is a first step towards a challenging task, and there are a number of limitations to our approach and possible scope for improvement.

\subsection{Internal Validity}
The biggest problem we faced during designing \textbf{BIMAN} was the lack of a golden dataset. We only knew about a handful of bots, which was not enough to design an accurate machine learning model. We tackled this problem by creating a dataset with one of the methods we proposed (\textbf{BIN}), and manually validating it. However, the bots found by \textbf{BIN} can have different characteristics than the rest of the bots, specifically, ones that may be trying to hide the fact that they are bots, and our method is not be able to detect them. 

We did not have a ground truth to validate the \textit{golden data} against (nor are we aware of the possibility of compiling such data with absolute certainty), so we had to come up with, what we judged to be, a reasonable alternative. The \textit{golden dataset} was manually curated by two authors of this paper, and the ambiguous cases, including the bots trying to disguise themselves as humans and vice versa, were excluded from the \textit{golden dataset}. 

Using the dataset generated by \textbf{BIN} as a golden dataset also means that we were not able to estimate the recall of \textbf{BIN} with it. Instead, we had to use a much smaller dataset to estimate its recall. Similarly, our final ensemble model was also trained with this smaller dataset, which led to some variation in its performance (AUC-ROC value varied between 0.89 and 0.95). 

Another threat to the effectiveness of our method is that a number of developers use automated scripts to handle some of their works, which uses their Git credentials while making commits. This is a tricky challenge for our method, since the signal from those authors appears mixed, and depending on what fraction of commits made using that author's ID is made by the bots, our method can fail to detect such authors as bots. Similarly, a few organizational IDs are sometimes used by bots as well as humans, and we have a similar issue regarding those IDs as well. We did not address the problem of multiple IDs belonging to the same author, however, we are testing different approaches for addressing this issue~\cite{fry2020dataset}, and, as a future work, plan on extending \textbf{BIMAN} with this capability. 

Provided that an estimated 1\% of the commit authors were found to be bots (Section~\ref{ss:estim}), an author detected as a bot by a 90\% accurate model has only only 8.3\% chance of actually being a bot (using Bayes' Theorem), i.e., we are bound to have false positives.


\subsection{Construct Validity}

The construct validity threats primarily apply to the \textbf{BIM} approach we used, since it was designed with specific ideas about how a bot might work.
\textbf{BIM} focuses on identifying bots that authored all the commit messages it is associated with, independent of whether they were generated by a template-based approach or not.
However, many developers make use of bots for generating certain commit messages (re-using the same author ID) and this hybrid classification is not addressed in this work.
The main factors that give rise to limitations are the content of commit messages,  number of commit messages per author, and performance of similarity measure.

The performance of \textbf{BIM} depends primarily on the performance of the similarity measure used to compare commit messages.
Commit messages tend to be concise, making it difficult to extract content characteristics (structural or semantic) that are useful for text similarity metrics.
Humans with consistent message styles become difficult to differentiate from template-based bot messages.
Moreover, if there are few commit messages or many unique messages, the document template score will not be effective, thus, this approach works better when enough data is available to almost saturate the document template score.
We note that \textbf{BIM}'s performance will also vary based on the language of the commit messages (e.g., Spanish and Chinese), and does not support multilingual sets of commit messages. 
\textbf{BIM}'s performance can be improved by using more effective similarity measures based on natural language processing~\cite{guzman2014sentiment}, document embeddings, clustering, and machine learning models~\cite{efthimion2018supervised}.


\subsection{External Validity} Our goal in this paper was to identify bots that make commits in social coding platforms. To that effect, our method of detecting bots could work for detecting other types of bots, such as pull-request bots, and chat bots. 

\section{Conclusion}\label{s:conclusion}
The automated nature of bots can inflate estimates of the amount of productivity and team size in software projects. 
Such bias may invalidate analyses and decisions based on these measures. Furthermore, bot activity may bias the estimates 
of social networks linking developers with bots or with other developers with whom they are not in contact. Bots, therefore, 
should be excluded from studies that focus on modeling the behavior of human developers in OSS projects. In this work we presented a novel approach, \textbf{BIMAN}, to detect bots using information from code commits. Our approach combines three independent models based on pattern matching, document similarity, and random forest classification. A significant portion of authors can be identified as bots using our proposed method, which can make studies of developers based on code commit data more accurate.

\section*{Acknowledgement}
The work has been partially supported by the following NSF awards:
CNS-1925615, IIS-1633437, and IIS-1901102. Vasilescu has been supported in part
by the NSF awards 1717415, 1901311.

\balance
\bibliographystyle{ACM-Reference-Format}
\bibliography{sigproc} 

\end{document}